\documentclass[]{elsarticle}

\usepackage{lineno,hyperref}
\usepackage{amsmath}
\modulolinenumbers[5]
\usepackage{subfigure}
\usepackage{xcolor}
\usepackage{siunitx}
\usepackage{soul}

\usepackage{paralist}
\usepackage{enumitem}

\DeclareSIUnit\sq{sq}

\journal{Journal}

\graphicspath{{figs/}}

\begin{document}

\begin{frontmatter}

\title{Kombucha electronics}

\author[1,4]{Andrew Adamatzky}
\author[2]{Giuseppe Tarabella}
\author[1]{Neil Phillips}
\author[3,1]{Alessandro Chiolerio}
\author[2]{Passquale D'Angelo}
\author[1]{Anna Nicolaidou}
\author[4,1]{Georgios~Ch.~Sirakoulis}
\address[1]{Unconventional Computing Laboratory, University of the West of England, Bristol, UK}
\address[2]{Institute of Materials for Electronic and Magnetism, National Research Council (IMEM-CNR), Parma (Italy)}
\address[3]{Istituto Italiano di Tecnologia, Center for Converging Technologies, Soft Bioinspired Robotics, Via Morego 30, 16165 Genova, IT}
\address[4]{Department of Electrical and Computer Engineering, Democritus University of Thrace, Xanthi, Greece}

\begin{abstract}
A kombucha is a tea and sugar fermented by over sixty kinds of yeasts and bacteria. This symbiotic community produces kombucha mats, which are cellulose-based hydrogels. The kombucha mats can be used as an alternative to animal leather in industry and fashion once they have been dried and cured. Prior to this study, we demonstrated that living kombucha mats display dynamic electrical activity and distinct stimulating responses. For use in organic textiles, cured mats of kombucha are inert. To make kombucha wearables functional, it is necessary to incorporate electrical circuits. We demonstrate that creating electrical conductors on kombucha mats is possible. After repeated bending and stretching, the circuits maintain their functionality. In addition, the abilities and electronic properties of the proposed kombucha, such as being lighter, less expensive, and more flexible than conventional electronic systems, pave the way for their use in a diverse range of applications.
\end{abstract}


\end{frontmatter}


\section{Introduction}

Kombucha is fermented by a symbiotic community of bacteria and yeasts~\cite{may2019kombucha,coelho2020kombucha,teoh2004yeast,may2019kombucha,kurtzman2001zygosaccharomyces,jarrell2000kombucha}. The symbiotic culture of bacteria and yeasts produces a cellulose-based hydro-gel, also known as bacterial cellulose, biofilm, commensal biomass, tea-fungus, scoby and zooglea. A tea fermented by the symbiotic community allegedly exhibits a range of health beneficial properties~\cite{vargas2021health,ivanivsova2020evaluation,coelho2020kombucha}, however these will not be discussed in the present work. 

Kombucha mats are unique symbiotic systems where over sixty species of yeasts and bacteria cooperate~\cite{may2019kombucha}. A kombucha is an example of a proto-multicellularity --- an organism combined of multiple species each one pursuing a common goal of prolonging a life time of the collective organism. Electrical properties of kombucha mats can further advance ideas on electricity based integration, and possibly, protocognition of symbiotic organisms~\cite{levin2012molecular,levin2014molecular, levin2019computational,levin2021bioelectric}. Similar bacterial cellulose mats, for example, produced by \textit{Acetobacter aceti} colonies, have been shown to feature interesting electrical properties and pressure sensing capabilities \cite{chiolerio2021acetobacter}.
  
Kombucha mats, when properly cured, show properties similar to textiles~\cite{wood2017microbes,laavanya2021current,domskiene2019kombucha,betlej2020influence,kaminski2020hydrogel,minh2021vegan}, and might make a competitive alternative to fungal leather and wearables~\cite{manan2022applications,gandia2021flexible}.

In a light of ongoing research on sensing and computing mechanisms embedded in living wearables~\cite{adamatzky2021towards,adamatzky2021reactive,chiolerio2022living,nikolaidou2022reactive} we aim to evaluate kombucha zoogleal mats as potentially embeddable cyber-physical wearable devices with non-linear and non-trivial electrical properties. To achieve the aim we test if basic components of the electrical circuits could made on dry kombucha mats.  

Modern electrical circuits require reliable electrical connections between electronic components (including sensors) and external signals for their construction and continued operation~\cite{whitaker2018electronics,wilamowski2018fundamentals,maini2018handbook}. Printed circuit boards (PCBs) are typically constructed from silkscreen, solder mask, copper, and substrate~\cite{jillek2005embedded,zheng2011review}. Material selection is crucial to the successful operation of printed circuit boards, especially thermal behaviour. The majority of PCB substrates fall into one of two categories: hard/rigid or soft/flexible. Ceramic-based materials typically provide excellent thermal conductivity, good dielectric properties, a high operating temperature, and a low expansion coefficient. The most popular rigid material is FR-4, a glass-reinforced epoxy laminate that is both inexpensive and versatile~\cite{mumby1989overview,ehrler2002properties}. Above a few GHz, the substantial dielectric loss (dissipation factor) of FR-4 renders it unsuitable for high-speed digital or high-frequency analogue circuits~\cite{mumby1989dielectric,djordjevic2001wideband}.

PCBs for wearables must often be mechanically flexible, waterproof, and shockproof and by default light~\cite{liu2022printed,kao2018skinwire,tao2017make,vieroth2009stretchable,buechley2009fabric}. Traditionally they are plastic based although they typically lack sustainability and cost-effectiveness. Polymeric soft materials offer superior resistance to stretching, bending and washing cycles~\cite{stoppa2014}. Moreover, wearables are intended to interact closely with their wearer, therefore bio-compatibility is advantageous, or at least resistance to the active chemical environment offered by the human skin. Therefore, the combination of bio-based PCBs and biodegradable components (including ICs) is especially advantageous for wearables.

\section{Methods and materials}
\label{methods}

\begin{figure}[!tbp]
    \centering
    \subfigure[]{\includegraphics[width=0.7\textwidth]{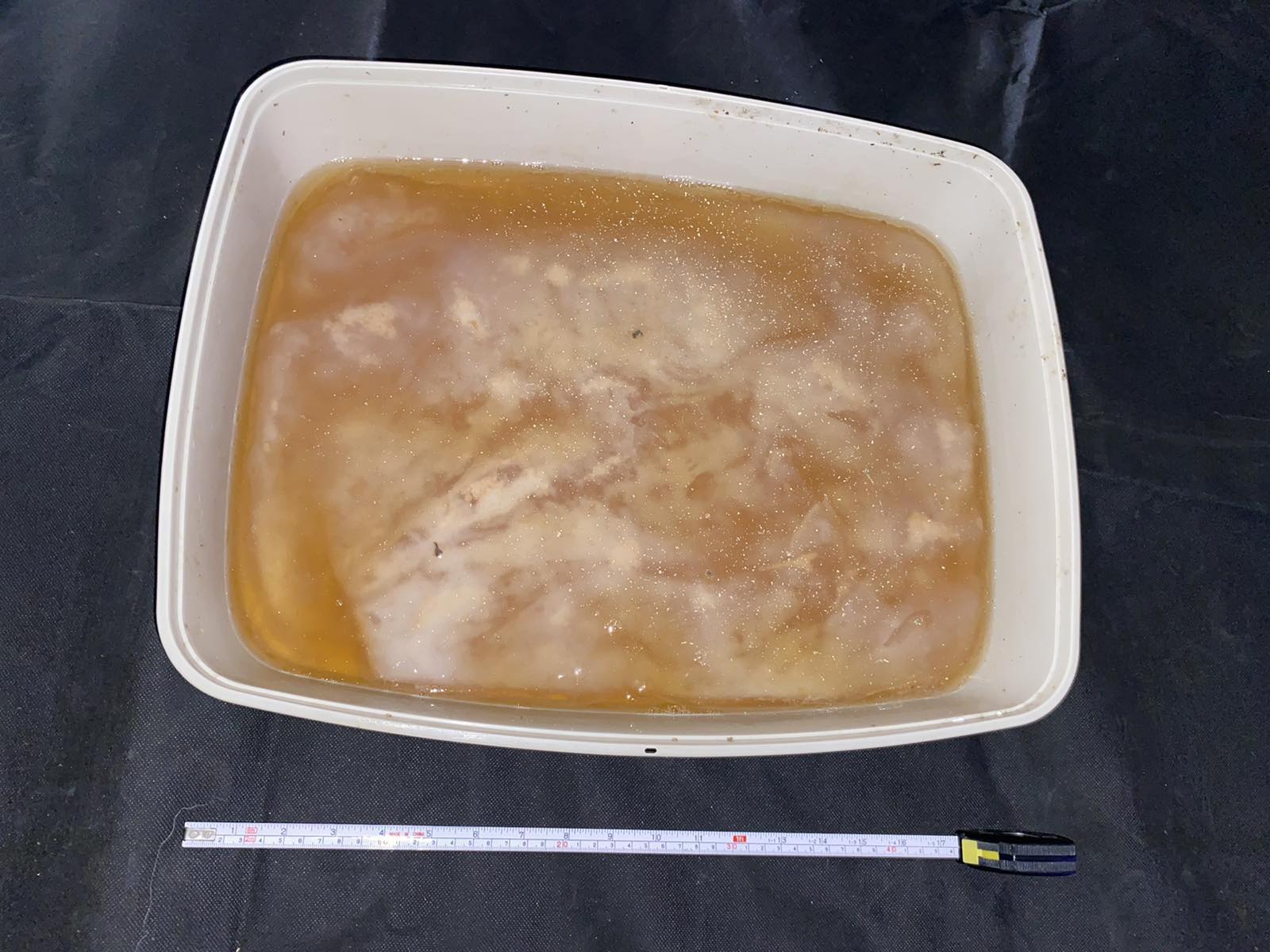}}
    \subfigure[]{\includegraphics[width=0.4\textwidth]{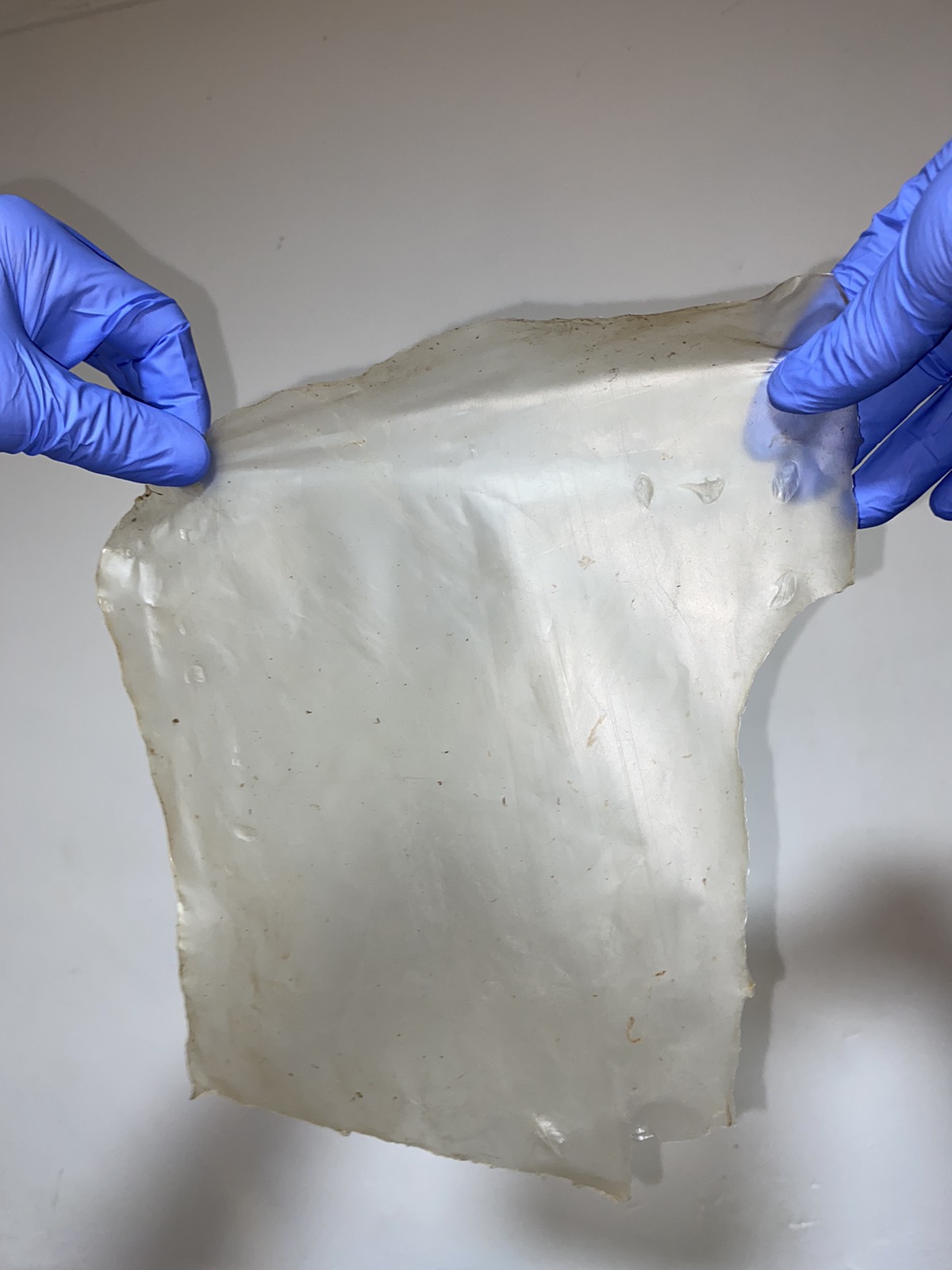} \, \includegraphics[width=0.4\textwidth]{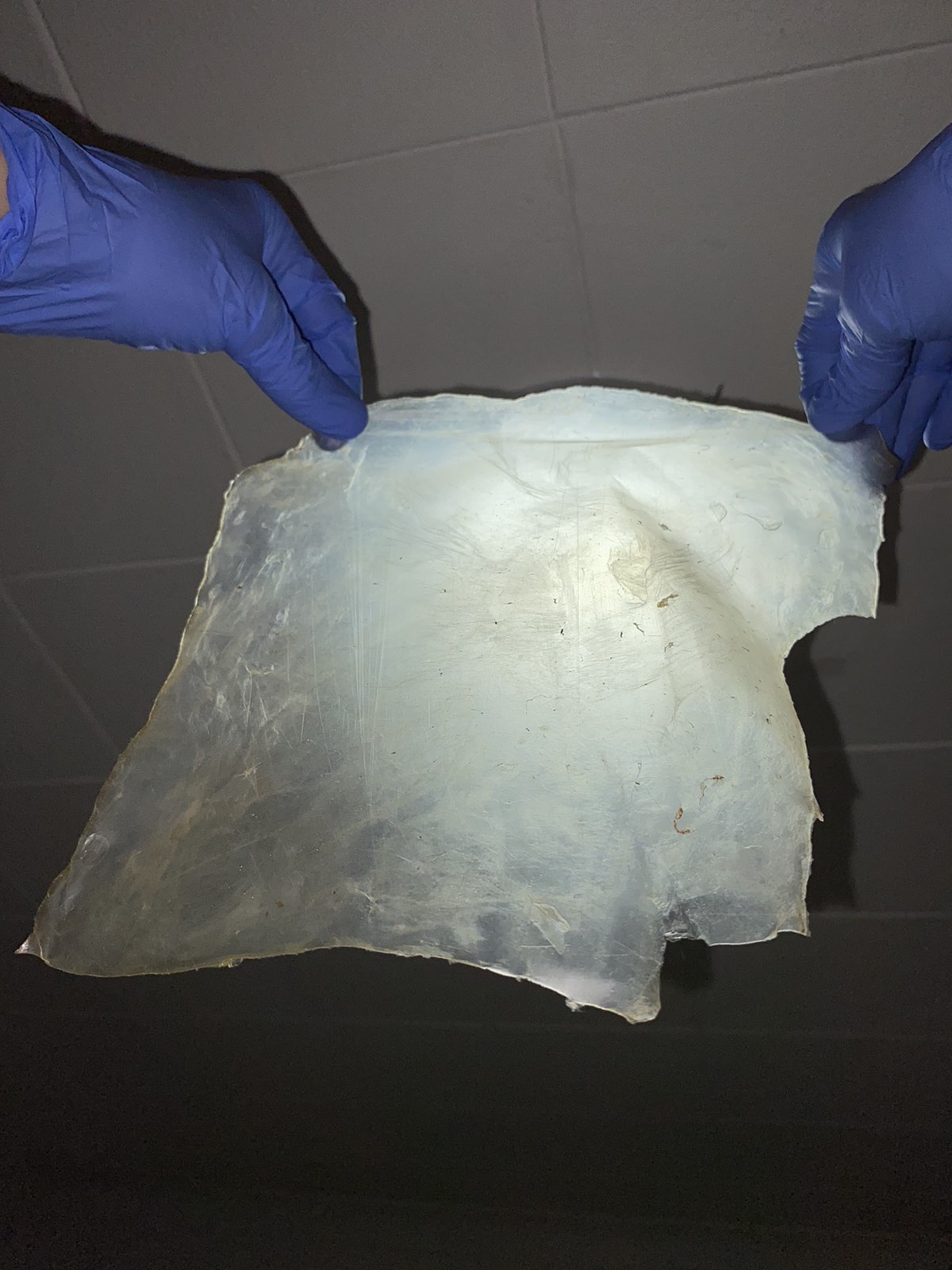}}
    \caption{(a)~Container with kombucha live mat on the surface of the liquid culture. (b)~Dried mat.}
    \label{fig:kombuchacontainer}
\end{figure}

The kombucha zooglea was commercially sourced (Freshly Fermented Ltd, UK) to grow mats of kombucha \emph{in situ}. The infusion was prepared as follows; 2\% tea (PG Tips, UK), 5\% sugar (Silver Spoon, UK), and \SI{1} {\litre} of boiled mains water. Containers with kombucha (Fig.~\ref{fig:kombuchacontainer}) were stored at ambient temperature (\SIrange{20}{23}{\celsius}) in darkness. The solution was refreshed each week. Kombucha mats were removed from the cultivation container and air-dried on plastic or paper at ambient temperature (several techniques were tried).

Four manufacturing technologies to add conductive tracks, attaching electronic components, and cut profiles of kombucha mats were explored.

Aerosol jet printing of PODOT:PSS was implemted as follows. 
Organic-based electrodes and interconnecting lines were printed by Aerosol Jet Printing (AJP200, Optomec, US \cite{aerosol}) by using an a highly-conductive inkjet formulation of PEDOT:PSS (Clevios P JET N V2, Heraeus, US \cite{heraeus}). Printing parameters were optimised for achieving conductive traces over the surface of kombucha mats used as the substrate. Electrochemical measurements were performed by a potentiostat (PalmSens4, PalmSens BV, NL \cite{palmsens}).

To 3D print TPU with 15\% carbon infill and metal-polymer composite --- biodegradable polyester and copper --- two compositions of filament (\SI{2.85} {\mm} diameter) were hot extruded onto kombucha mats via \SI{0.4} {\mm} nozzle on 3D printer (S5, Ultimaker, UK \cite{S5}). The composition filaments were `Conductive Filaflex Black' rated \SI{3.9} {\ohm\per\cm} \cite{filaflex} and `Electrifi Conductive Filament' rated \SI{0.006} {\ohm\per\cm} \cite{electrifi}.

Conductive pathways were drawn onto kombucha mats with two compositions of conductive ink, including `Bare Conductive' rated \SI{55} {\ohm\per\sq} at \SI{50}{\micro\meter} thickness \cite{bareconductive} and `XD-120 conductive silver ink' rated rated \SI{0.00003} {\ohm\per\cm} \cite{silver}. 

When shaping was involved, kombucha mats of 0.45\SI{\pm 0.1} {\mm} thickness were cut with \SI{75} {\watt} CNC laser cutter (Legend 36EXT, Epiloglasers, US, \cite{Legend}) while parameters (speed of motion, beam power, pulses per inch) were adjusted to determine the optimal settings.

\section{Results}

Kombucha mats are proven to be robust to tearing, and are not destroyed even by immersion in water for several days. The mat survived oven temperature up to 200C but burn when exposed to open flame. 
We have demonstrated that it possible to (1)~precisely cut kombucha mats with laser, (2)~aerosol jet print PODOT:PSS circuits on kombucha mats, (3)~3D printing TPU and metal-polymer composite on kombucha mats, (4)~draw conductive tracks and arrange functional elements with conductive paints.

\begin{figure}[!tbp]
    \centering
    \includegraphics[width=0.7\textwidth]{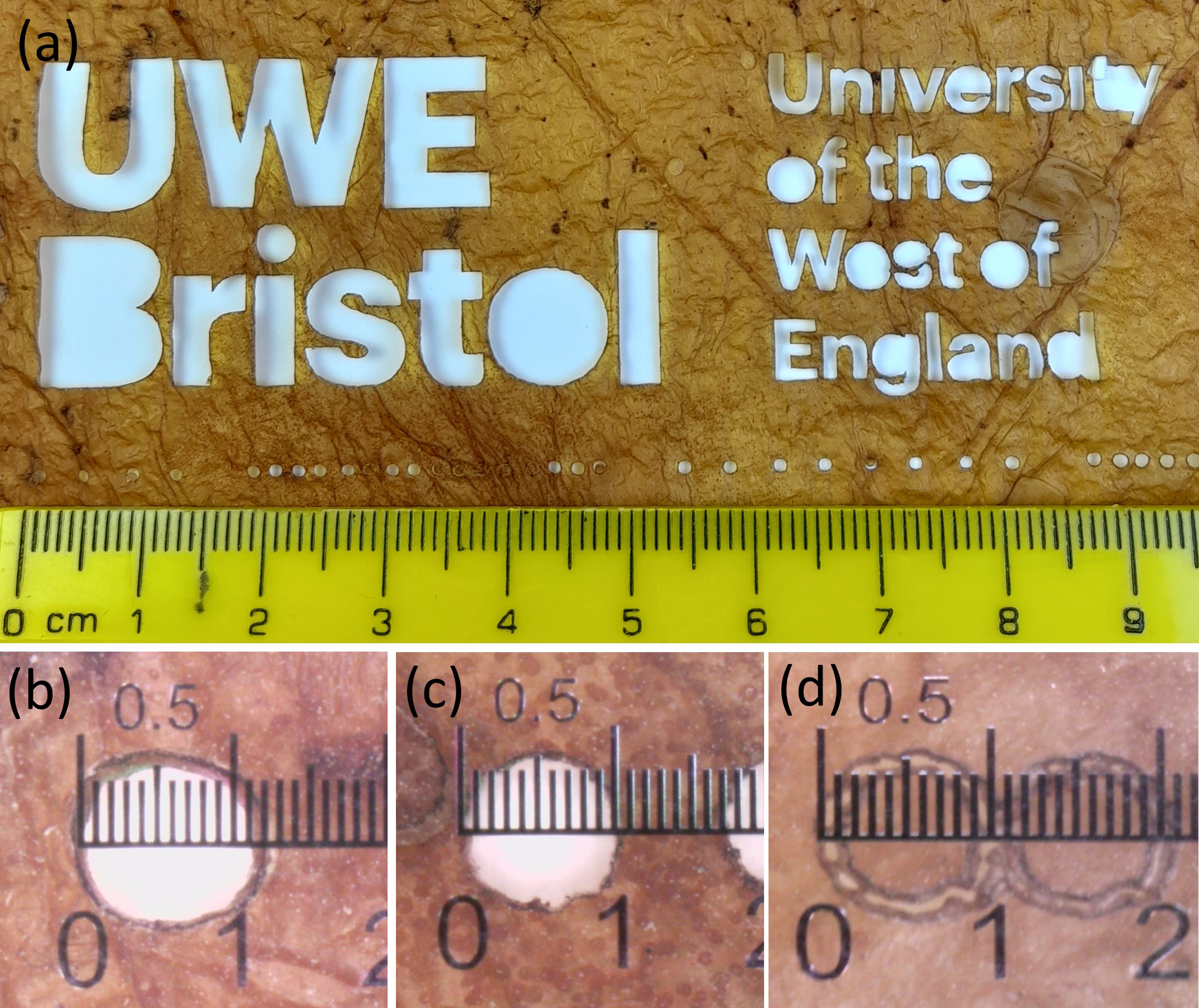}
    \caption{Kombucha mats cut with laser cutter 
    (a)~letters and holes of different sizes, 
    (b)~nominal \SI{1} {\mm} hole cut to $\sim$\SI{1.1} {\mm} diameter with $\sim$\SI{25} {\watt} laser power, 
    (c)~nominal \SI{1} {\mm} hole cut to $\sim$\SI{1.0} {\mm} diameter with $\sim$\SI{18} {\watt} laser power, 
    (d)~nominal \SI{1} {\mm} holes only partly cut out to $\sim$\SI{1.0} {\mm} diameter with $\sim$\SI{10} {\watt} laser power.
    }
    \label{fig:laser}
\end{figure}

Laser cutting proved to be a problem free procedure. Exemplars of kombucha mats cut with laser cutter are shown in Fig.~\ref{fig:laser}. The laser settings (e.g. speed of motion, beam power and number of laser pulses per inch) were found to be critical to accurate cutting. The optimal setting for 0.45\SI{\pm 0.1} {\mm} thickness was found to be 80 inches per second, $\sim$\SI{18} {\watt}, and 500 pulses per inch, as shown in Fig.~\ref{fig:laser}(c). If the beam power is raised above optimal level the cut becomes wider than desirable, as shown in Fig.~\ref{fig:laser}(b). Conversely, if the beam power is lower than optimal level the mat is only partly cut through, as shown in Fig.~\ref{fig:laser}(d). With optimised settings, kombucha mats were found to cut well with minimal smoke. Some cut sections needed to be agitated free for removal.

Organic electrical conductors have been printed by Aerosol Jet Printing (AJP) with the aim of creating circuits over kombucha mats, exploited as potential substrates in wearable electronics. Circuits over kombucha can act in perspective as sensors or biosensors, coupled also with printed antennas for wireless data communication and storing in clouds. Herein, we are going to explore basic properties of printed traces over the surface of kombucha.

Aerosol Jet Printing is particularly suitable for printing over irregular surfaces, flexible and/or stretchable substrates made of natural materials (bio-polymers) because of it operates in non-contact mode at a fixed distance from the substrate. Basic principles and mechanisms of AJP techniques have been discussed in literature \cite{Tarabella_2020,secor2018principles,wilkinson2019review,mette2007metal}. This technology belongs to the additive manufacturing sector and offers advantages with respect to other well known technologies and broadly distributed, such as ink jet printing (normally referred to liquid inks jetted using thermal or piezoelectric nozzles \cite{scalisi2015}).

A highly conductive formulation of PEDOT:PSS was used as ink: 2mL of ink was uploaded in the ultrasonic atomiser of AJP 200, by setting the gas flows at 30 and 25 sccm for the atomiser and sheath gas, respectively. A 200um size nozzle was mounted on the printed head. The printing run was operated on cool-conditions to avoid exposure of kombucha to heat treatments. Elementary circuit elements were printed firstly, 3 circular electrodes (2mm diameter) at a fixed distance, acting as the working, counter and reference electrodes, for the evaluation of impedance of the electrode-kombucha interface, by Impedance Electrochemical Spectroscopy (EIS) analysis.

\begin{figure}[!tbp]
    \centering
\subfigure[]{\includegraphics[width=0.38\textwidth]{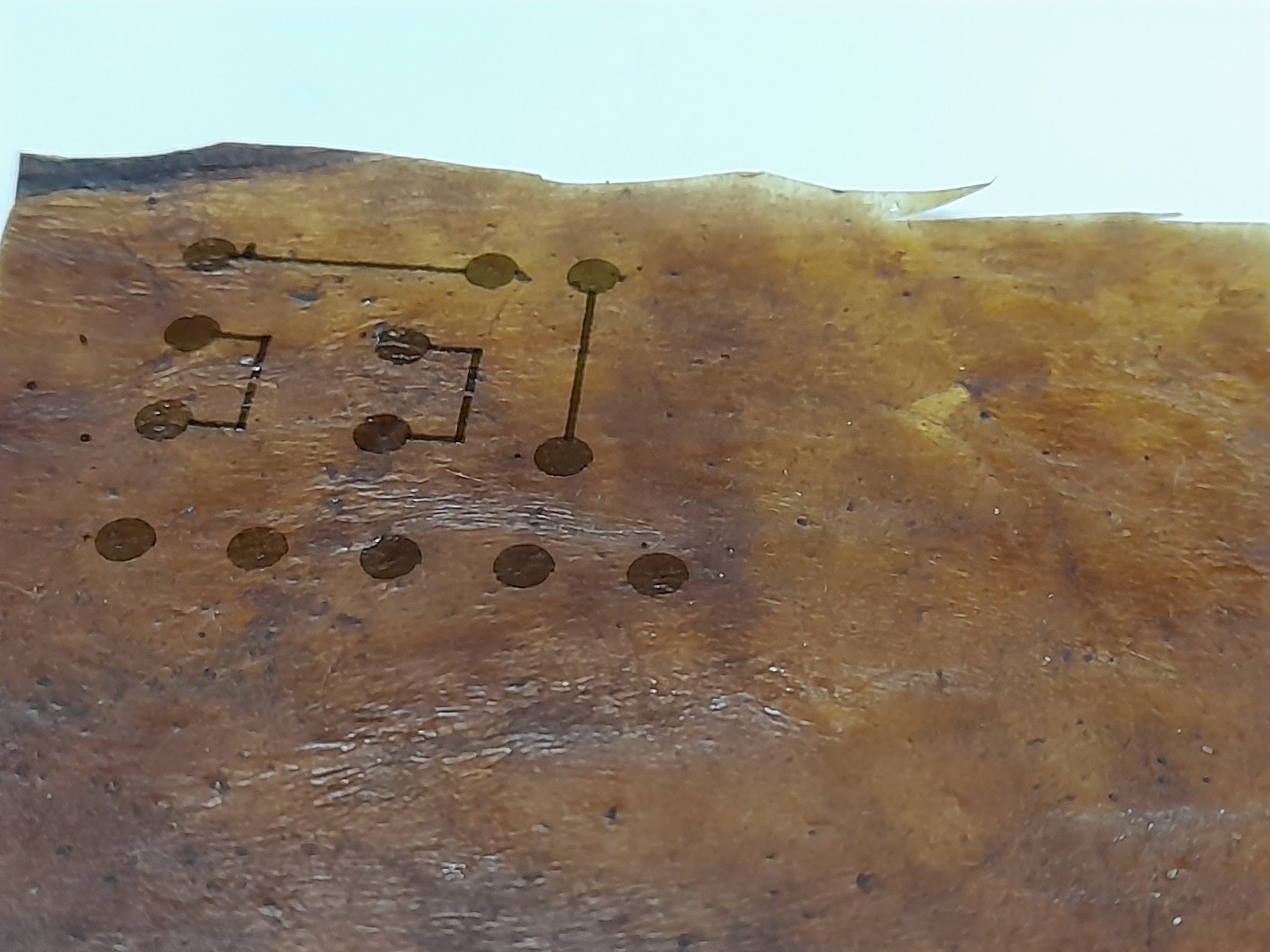}}
\subfigure[]{\includegraphics[width=0.38\textwidth]{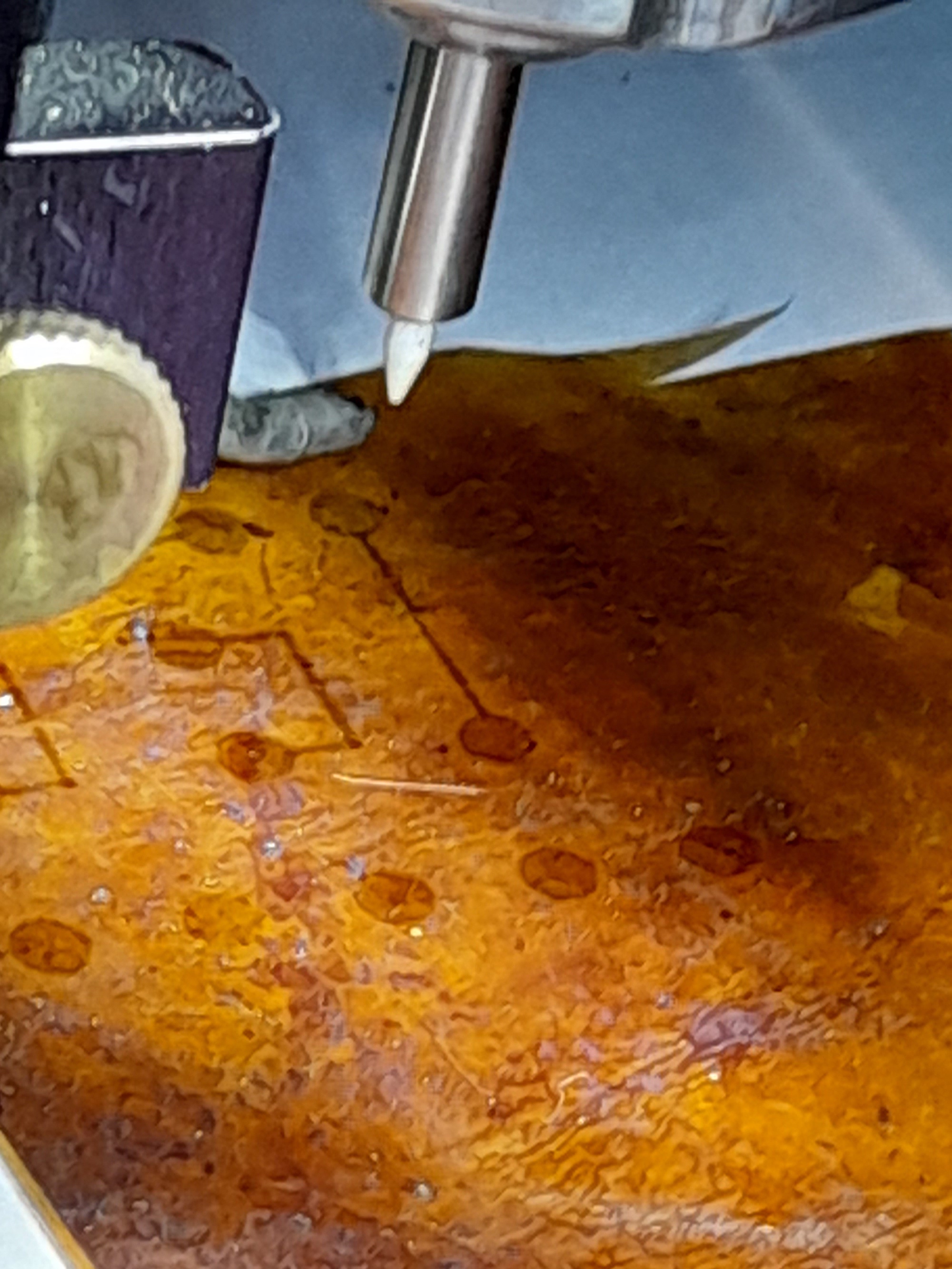}}
\subfigure[]{\includegraphics[width=0.38\textwidth]{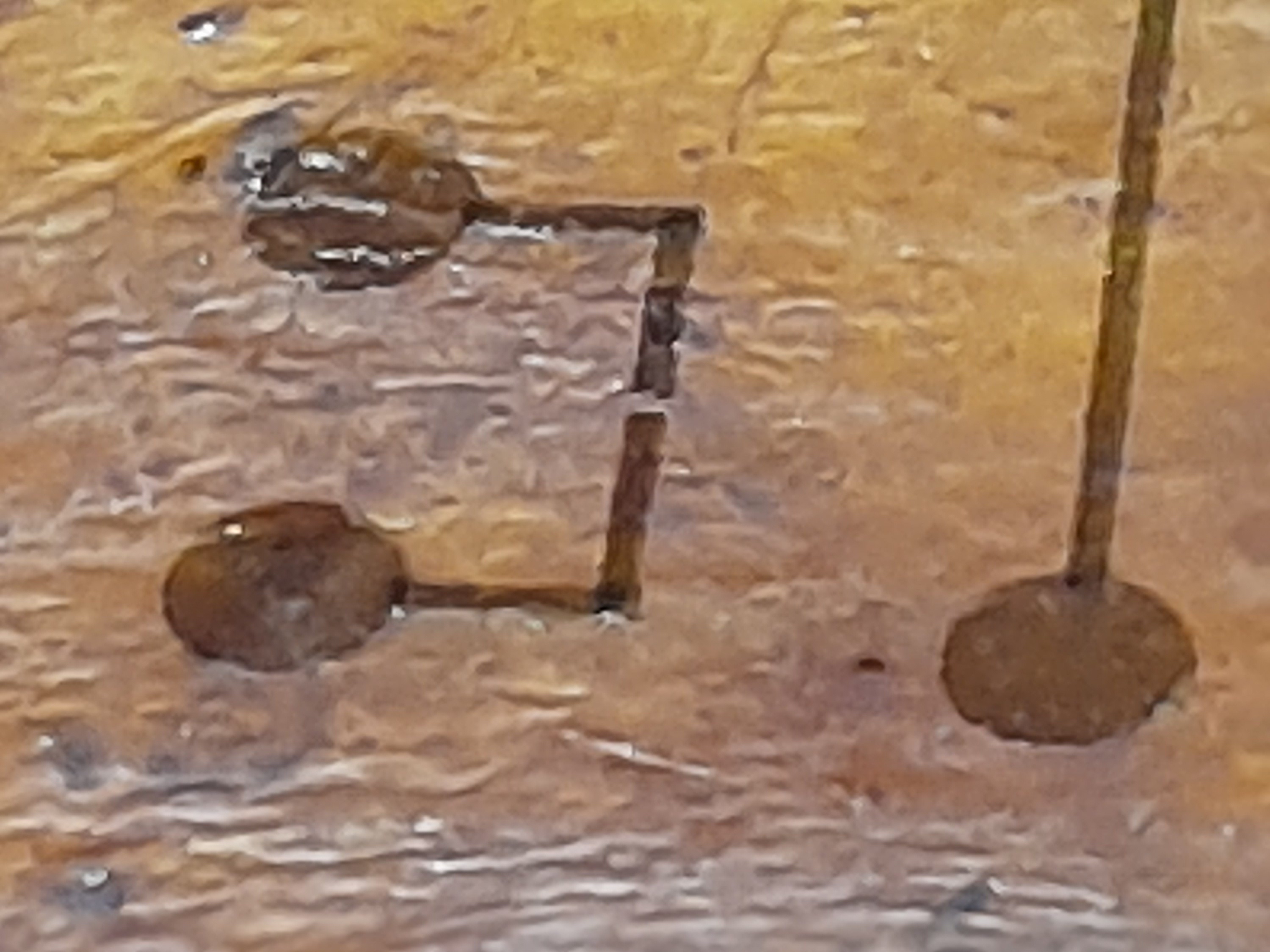}}
\subfigure[]{\includegraphics[width=0.38\textwidth]{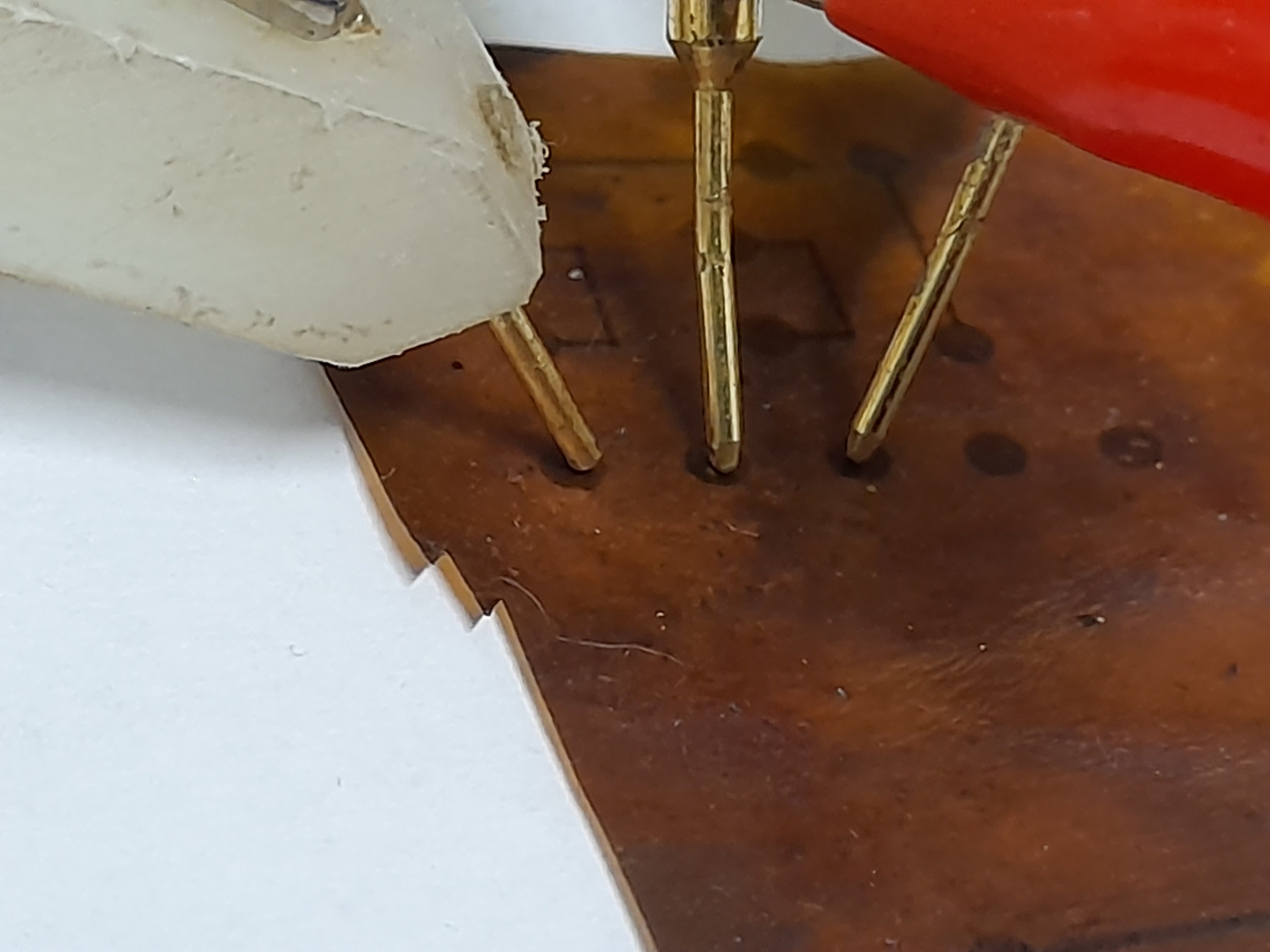}}
\subfigure[]{\includegraphics[width=0.38\textwidth]{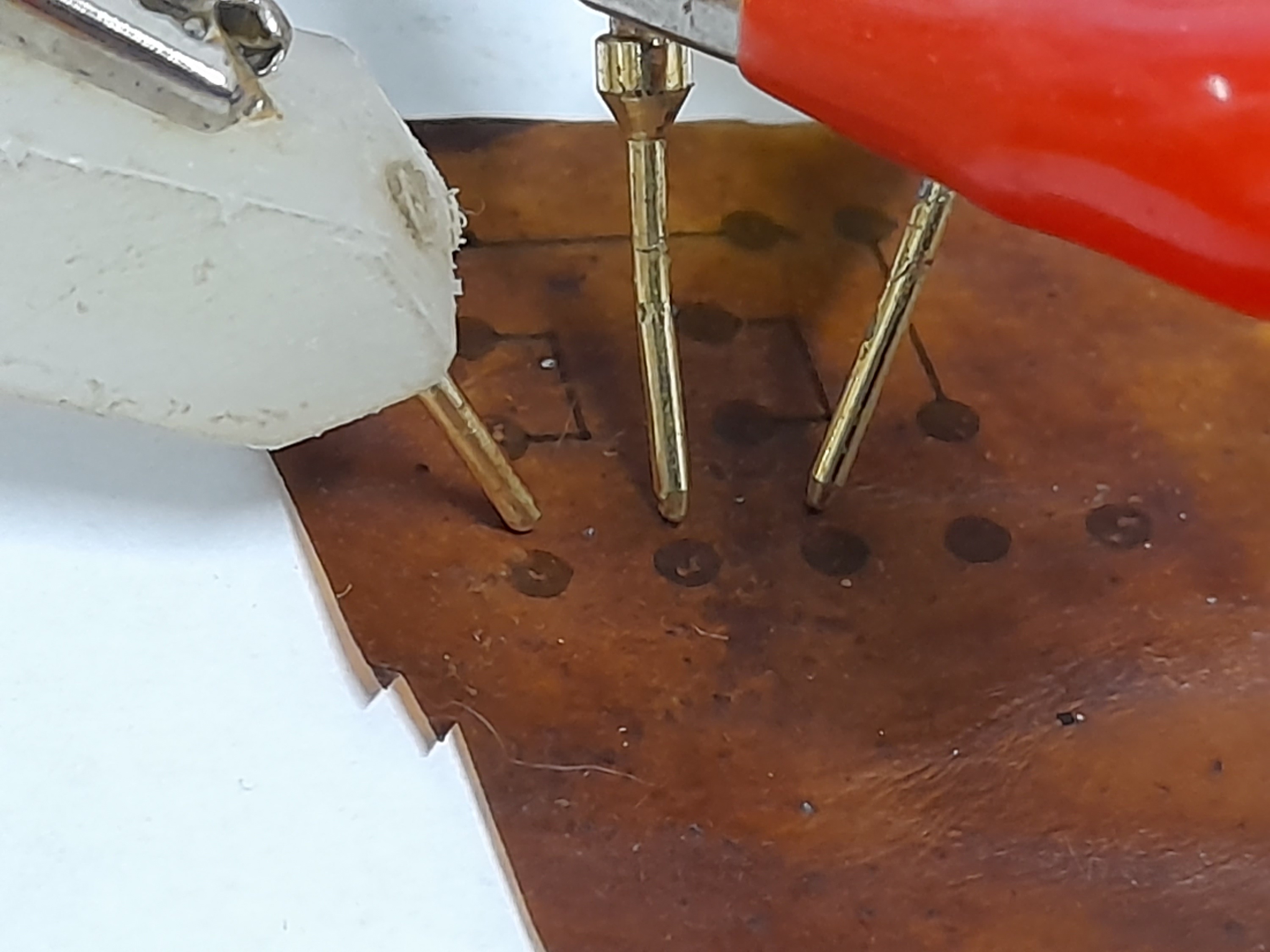}}
\subfigure[]{\includegraphics[width=0.38\textwidth]{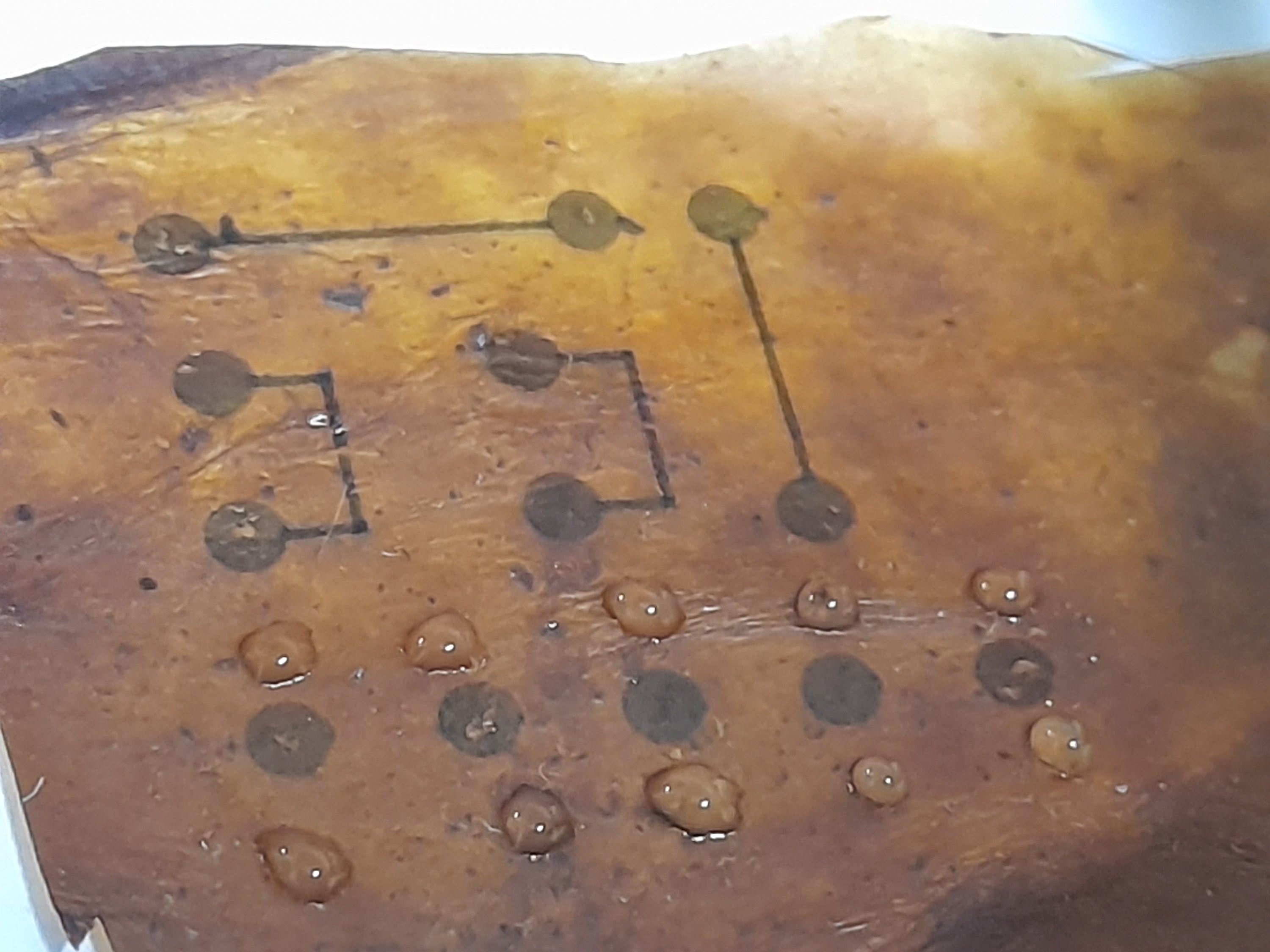}}
    \caption{Exemplars of deposition of PEDOT:PSS circuits and measurements of electrical properties (a) PEDOT:PSS round pads at a fixed distance from each other with interconnecting tracks (b) Aerosol Jet Printing nozzle (c) defined gap between tracks (d) spring loaded electrodes on PEDOT:PSS pads (e) spring loaded electrodes on surface of kombucha (f) hydration of PEDOT:PSS 
    }
    \label{fig:pedotexamples}
\end{figure}

\begin{figure}[!tbp]
    \centering
\subfigure[]{\includegraphics[width=0.69\textwidth]{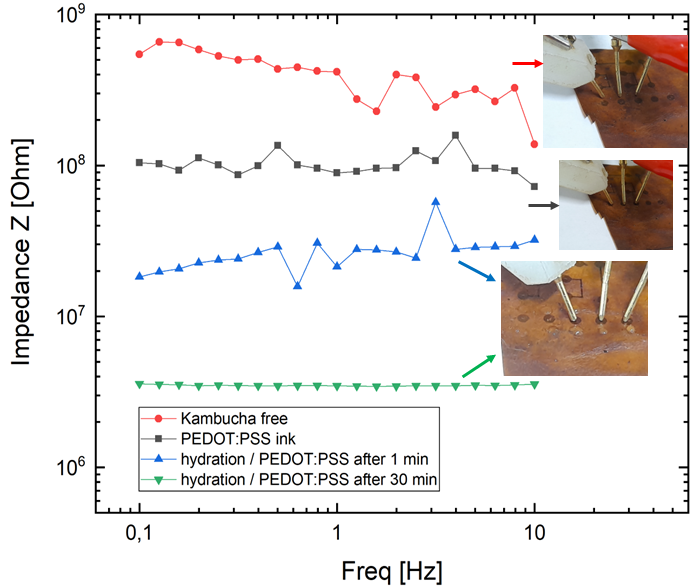}}
\subfigure[]{\includegraphics[width=0.3\textwidth]{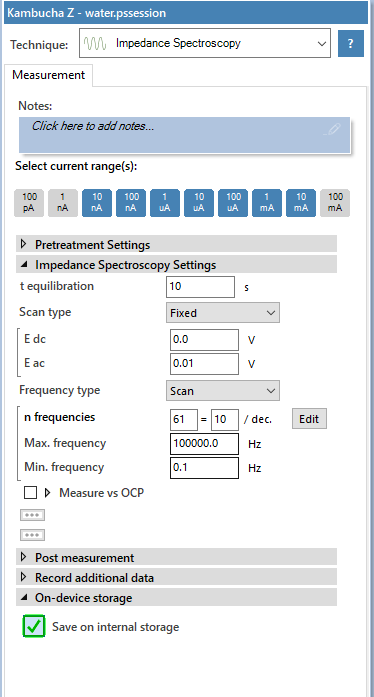}}
    \caption{Electrical properties of kombucha mat with and without PEDOT:PSS circuits (a) impedance against frequency (b) spectroscopy settings.}
    \label{fig:pedoperformance}
\end{figure}

Exemplars of deposition of PEDOT:PSS circuits and measurements of electrical properties are shown in Fig.~\ref{fig:pedotexamples}.
The figure shows the acquired data of EIS over 1) three free points over the kombucha surface; 2) three PEDOT:PSS electrodes used as the working (RE), counter (CE) and reference (RE) electrodes, placed at fixed distances and the same as the free points of 1); 3) the same measurements of 2) after hydration, where hydration was performed by placing 20 $\mu L$ of water drops in the surrounding area of the electrodes  over the kombucha surface. Being kombucha a cellulose-based material, it is very sensitive to water absorption, and up-taking water in the kombucha backbone make the kombucha foil more conductive. Impedance measurements increase almost instantly after water dropping, and stabilise quickly; the measurements after 30 min after water dropping show a more stable signal.  Electrical properties of kombucha mat with and without PEDOT:PSS circuits are shown in Fig.~\ref{fig:pedoperformance}.

\begin{figure}[!tbp]
    \centering
    \includegraphics[width=0.6\textwidth]{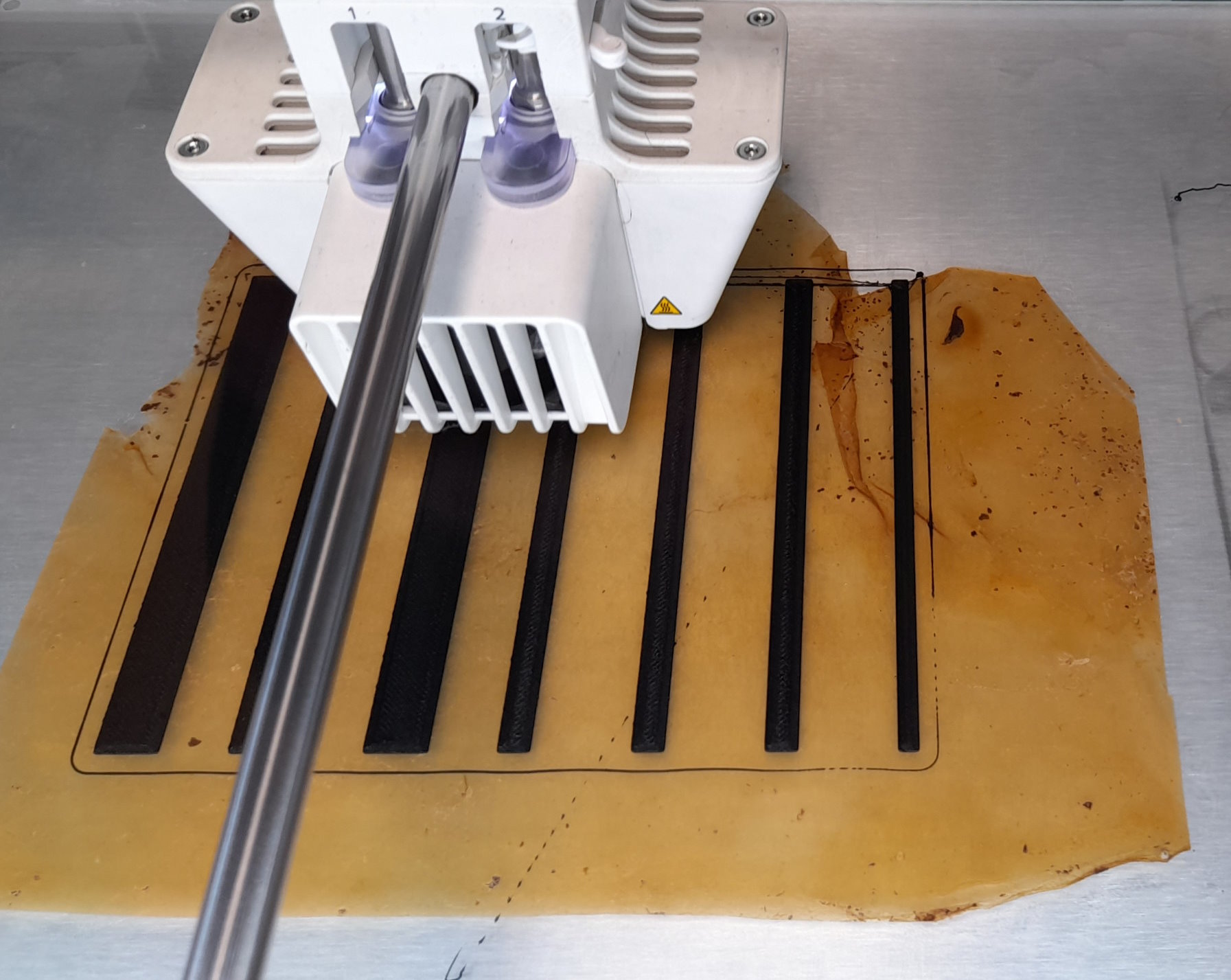}
    \caption{3D printing of flexible TPU (with 15\% carbon infill) tracks on kombucha mat.}
    \label{fig:TPU}
\end{figure}

Exemplar of tracks of TPU (with 15\% carbon infill) 3D printed on kombucha mat are shown in Fig.~\ref{fig:TPU}. Track resistance of TPU (with 15\% carbon infill) and Electrifi (metal-polymer composite --- biodegradable polyester and copper) was found to vary with width and thickness,as summarised in Tab.~\ref{table:TPUresistance}. Tracks of \SI{100} {\mm} length were measured with LCR meter (891, BK Precision, UK). Flexibility of the tracks was found to vary with thickness. Both TPU and Electrifi tracks remained attached to kombucha after a couple of days of immersion in water. Their attachment might be 'mechanical' rather than chemical as the liquid (melted) polymers are effectively 'injected' into/onto the surface of kombucha effectively filling any surface irregularities which then act as 'grips' holding track in position.

\begin{table}[tbp]
    \centering
    \caption{Track resistance of TPU (with 15\% carbon infill) and Electrifi.}
    \begin{tabular}{|c|c|c|c|c|}
    \hline
        \textbf{Width} & \textbf{Thick-} & \textbf{Cross-} & \textbf{TPU} & \textbf{Electrifi} \\
         ~ & \textbf{ness} & \textbf{section} & with 15\% carbon & ~ \\ 
        {(mm)} & {(mm)} & {(mm$^2$)} & {(\SI{}{\ohm\per\centi\metre})} & {(\SI{}{\ohm\per\centi\metre})} \\ \hline
        10 & 5 & 50 & 650 & ~  \\ \hline
        10 & 3 & 30 & 920 & ~ \\ \hline
        10 & 1 & 10 & 2,450 & ~  \\ \hline
        5 & 3 & 15 & 1,400 & ~ \\ \hline
        5 & 2 & 10 & 2,120 & 0.28 \\ \hline
        5 & 1 & 5 & 3,440 & 0.43 \\ \hline
        3 & 1 & 3 & 4,490 & 0.89 \\ \hline
    \end{tabular}
    \label{table:TPUresistance}
\end{table}

\begin{figure}[htbp]
    \centering
\subfigure[]{\includegraphics[width=0.6\textwidth]{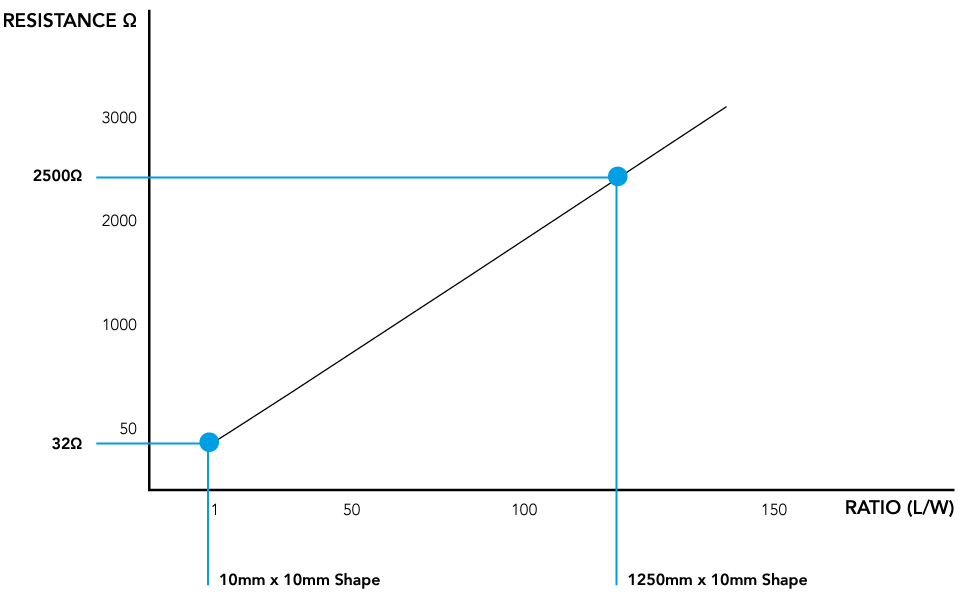}}
    \caption{Electrically conductive paint, data-set from Bare Conductive (UK).}
    \label{fig:examples}
\end{figure}

With regards to electrically conductive paint, experiments demonstrated that `Bare Conductive'~\cite{paintdata} adheres well to the kombucha mats and sustains some degree of flexibility. Typical electrical conductivity for tracks is shown in Fig.~\ref{fig:examples}. Track resistance of the conductive paint tracks on kombucha mats varied between \SIrange {20} {200} {\ohm\per\centi\metre}. These values roughly align with the `Bare Conductive' data-sheet~\cite{paintdata} with `thick' tracks. Track resistance of XD-120 conductive silver ink on kombucha mat was also found to vary. Typical range \SIrange {1.5} {10} {\ohm\per\centi\metre}

\section{Discussion}


Four technologies for manufacturing kombucha based PCBs were explored aerosol jet printing of PODOT:PSS, 3D printing of TPU and metal-polymer composite, adding ink with conductive filler and laser cutting. Each offered advantages and disadvantages compared to other technologies.

\begin{figure}[htbp]
    \centering
    \includegraphics[width=0.7\textwidth]{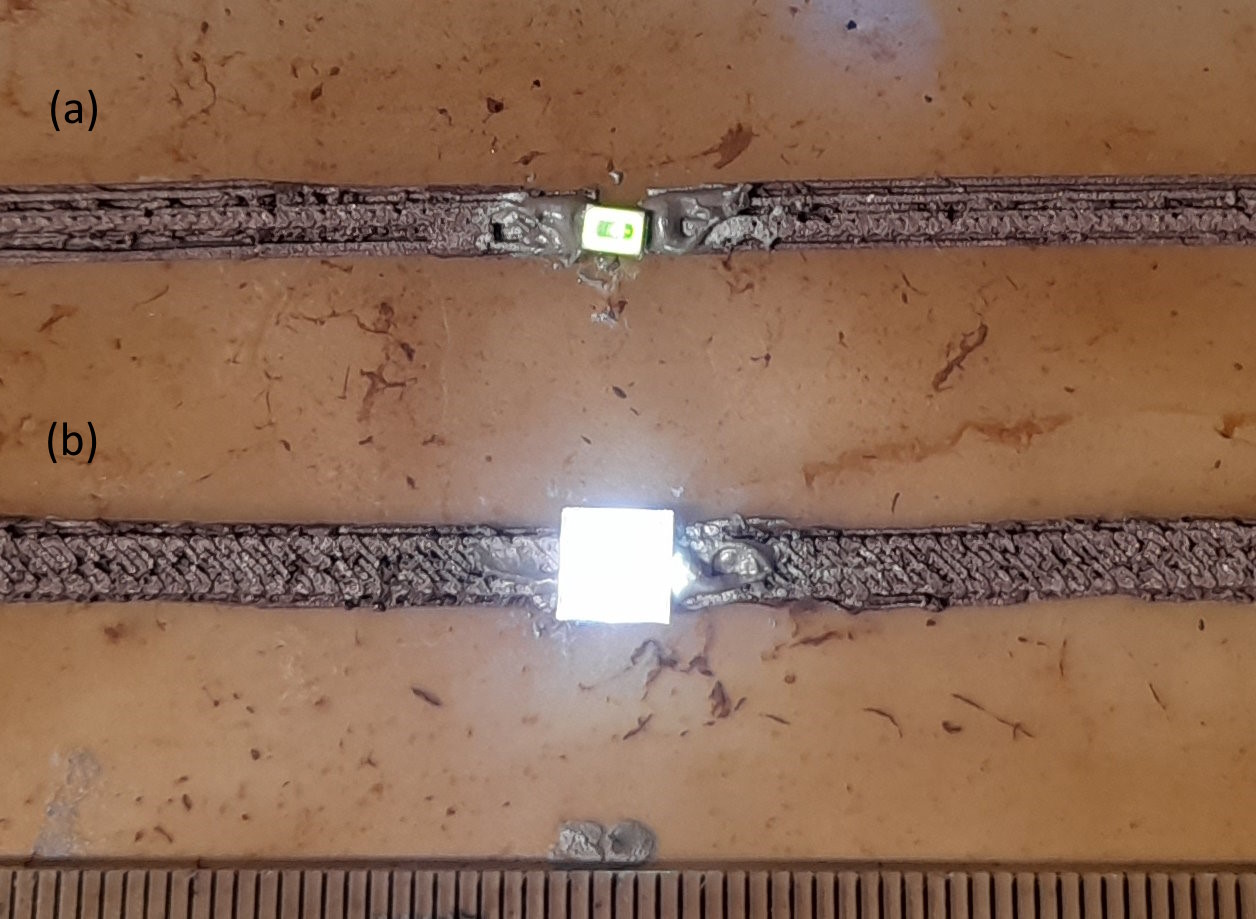}
    \caption{Exemplar of metal-polymer composite (Electrifi) tracks on kombucha mat (a) $\sim$\SI{3} {\mm} wide track with SMD LED (3020 package) green colour (b) $\sim$\SI{5} {\mm} wide track with SMD LED (5050 package) white colour (scale of ruler in mm).}
    \label{fig:LEDs}
\end{figure}

As demonstrated in Fig.~\ref{fig:LEDs}, it is feasible to construct electrical circuits on kombucha mats. Two track widths ($\sim$\SI{3}{} and $\sim$\SI{5} {\mm}) and two packages (3020 and 5050) of surface mount devices (SMD) are displayed. A silver-loaded, conductive, two-part epoxy (Chemtronics CW2400 \cite{CW2400}) was manually applied to mechanically attach and electrically connect SMDs to polymer tracks. For volume manufacture, SMDs would be automatically mounted using a pick and place machine and conductive epoxy precisely and automatically dispensed with in-line dispensers.

\begin{figure}[htbp]
    \centering
    \includegraphics[width=0.95\textwidth]{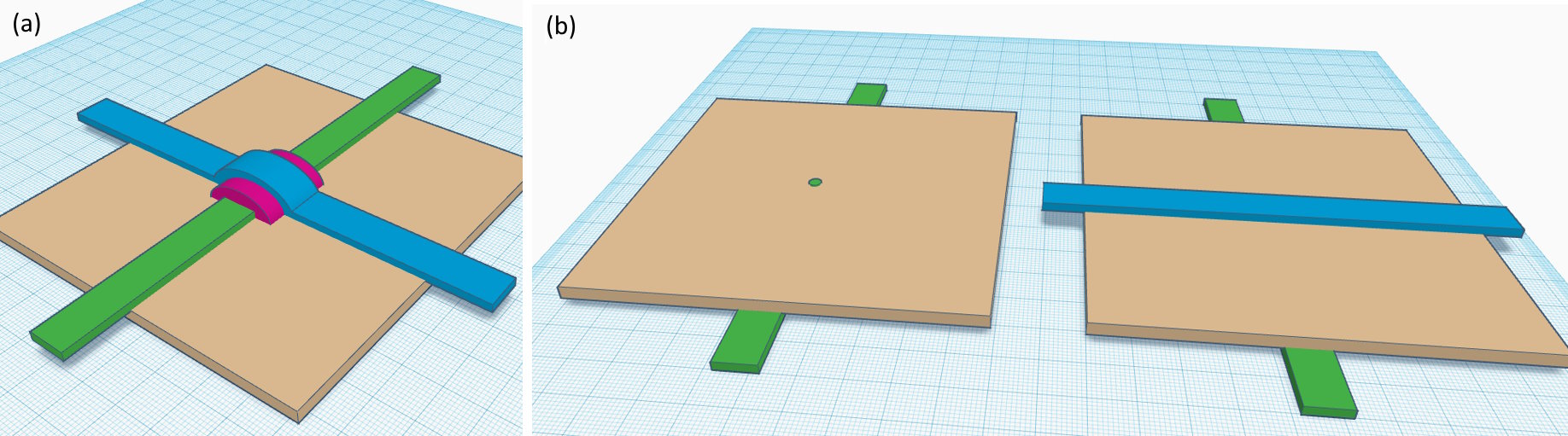}
    \caption{Methods of cross connecting on kombucha mats (a)~single sided cross-over bridge with insulator between (b)~through-hole double sided with laser hole cutting.}
    \label{fig:connections}
\end{figure}

Two potential methods of forming cross connections on kombucha mats via 3D printing of conductive material --- single sided cross-over bridges and through-hole double sided via laser hole cutting --- are illustrated  Fig.~\ref{fig:connections}.

Kombucha mats show properties that can be exploited to envision potential and future kombucha-based devices. The hydration-dependent electrical conduction of kombucha allows to extend the  potential operational frequency range of surface-electrodes over kombucha mats, as well as to exploit the kombucha mat as a resistive switching device in a planar electro-chemical cell. 

Future research will be concerned with printing advanced functional circuits, capable for detecting, and may be recognising, mechanical, optical and chemical stimuli, implementing sensorial fusion and distributed information processing. 

\section{Acknowledgements}
We are grateful to Geoff Sims for laser cutting of kombucha mats. We thank Ultimaker/MakerBot Technical Support for guidance on optimising print settings with Ultimaker S5. We thank Dr Shengrong Ye (Multi3D) for guidance on 3D printing Electrifi filament.

\section{Availability of data} 
The raw data-sets obtained in this study are available from the corresponding author on reasonable request.


\begin{thebibliography}{10}
\expandafter\ifx\csname url\endcsname\relax
  \def\url#1{\texttt{#1}}\fi
\expandafter\ifx\csname urlprefix\endcsname\relax\def\urlprefix{URL }\fi
\expandafter\ifx\csname href\endcsname\relax
  \def\href#1#2{#2} \def\path#1{#1}\fi

\bibitem{may2019kombucha}
A.~May, S.~Narayanan, J.~Alcock, A.~Varsani, C.~Maley, A.~Aktipis, Kombucha: a
  novel model system for cooperation and conflict in a complex multi-species
  microbial ecosystem, PeerJ 7 (2019) e7565.

\bibitem{coelho2020kombucha}
R.~M.~D. Coelho, A.~L. de~Almeida, R.~Q.~G. do~Amaral, R.~N. da~Mota, P.~H.~M.
  de~Sousa, Kombucha, International Journal of Gastronomy and Food Science 22
  (2020) 100272.

\bibitem{teoh2004yeast}
A.~L. Teoh, G.~Heard, J.~Cox, Yeast ecology of kombucha fermentation,
  International journal of food microbiology 95~(2) (2004) 119--126.

\bibitem{kurtzman2001zygosaccharomyces}
C.~P. Kurtzman, C.~J. Robnett, E.~Basehoar-Powers, Zygosaccharomyces
  kombuchaensis, a new ascosporogenous yeast from ‘kombucha tea’, FEMS
  Yeast Research 1~(2) (2001) 133--138.

\bibitem{jarrell2000kombucha}
J.~Jarrell, T.~Cal, J.~Bennett, The kombucha consortia of yeasts and bacteria,
  Mycologist 14~(4) (2000) 166--170.

\bibitem{vargas2021health}
B.~K. Vargas, M.~F. Fabricio, M.~A.~Z. Ayub, Health effects and probiotic and
  prebiotic potential of kombucha: A bibliometric and systematic review, Food
  Bioscience 44 (2021) 101332.

\bibitem{ivanivsova2020evaluation}
E.~Ivani{\v{s}}ov{\'a}, K.~Me{\v{n}}hartov{\'a}, M.~Terentjeva, L.~Harangozo,
  A.~K{\'a}ntor, M.~Ka{\v{c}}{\'a}niov{\'a}, The evaluation of chemical,
  antioxidant, antimicrobial and sensory properties of kombucha tea beverage,
  Journal of food science and technology 57~(5) (2020) 1840--1846.

\bibitem{levin2012molecular}
M.~Levin, Molecular bioelectricity in developmental biology: new tools and
  recent discoveries: control of cell behavior and pattern formation by
  transmembrane potential gradients, Bioessays 34~(3) (2012) 205--217.

\bibitem{levin2014molecular}
M.~Levin, Molecular bioelectricity: how endogenous voltage potentials control
  cell behavior and instruct pattern regulation in vivo, Molecular biology of
  the cell 25~(24) (2014) 3835--3850.

\bibitem{levin2019computational}
M.~Levin, The computational boundary of a “self”: developmental
  bioelectricity drives multicellularity and scale-free cognition, Frontiers in
  Psychology 10 (2019) 2688.

\bibitem{levin2021bioelectric}
M.~Levin, Bioelectric signaling: Reprogrammable circuits underlying
  embryogenesis, regeneration, and cancer, Cell 184~(8) (2021) 1971--1989.

\bibitem{chiolerio2021acetobacter}
A.~Chiolerio, A.~Adamatzky, Acetobacter biofilm: Electronic characterization
  and reactive transduction of pressure, ACS Biomaterials Science \&
  Engineering 7~(4) (2021) 1651--1662.

\bibitem{wood2017microbes}
J.~Wood, Are microbes the future of fashion?, The Microbiologist 18~(2) (2017).

\bibitem{laavanya2021current}
D.~Laavanya, S.~Shirkole, P.~Balasubramanian, Current challenges, applications
  and future perspectives of scoby cellulose of kombucha fermentation, Journal
  of Cleaner Production 295 (2021) 126454.

\bibitem{domskiene2019kombucha}
J.~Domskiene, F.~Sederaviciute, J.~Simonaityte, Kombucha bacterial cellulose
  for sustainable fashion, International Journal of Clothing Science and
  Technology 31~(5) (2019) 644--652.

\bibitem{betlej2020influence}
I.~Betlej, R.~Salerno-Kochan, K.~J. Krajewski, J.~Zawadzki, P.~Boruszewski, The
  influence of culture medium components on the physical and mechanical
  properties of cellulose synthesized by kombucha microorganisms, BioResources
  15~(2) (2020) 3125--3135.

\bibitem{kaminski2020hydrogel}
K.~Kami{\'n}ski, M.~Jarosz, J.~Grudzie{\'n}, J.~Pawlik, F.~Zastawnik,
  P.~Pandyra, A.~M. Ko{\l}odziejczyk, Hydrogel bacterial cellulose: A path to
  improved materials for new eco-friendly textiles, Cellulose 27~(9) (2020)
  5353--5365.

\bibitem{minh2021vegan}
N.~T. Minh, H.~N. Ngan, Vegan leather: An eco-friendly material for sustainable
  fashion towards environmental awareness, in: AIP Conference Proceedings, Vol.
  2406, AIP Publishing LLC, 2021, p. 060019.

\bibitem{manan2022applications}
S.~Manan, O.~M. Atta, A.~Shahzad, M.~Ul-Islam, M.~W. Ullah, G.~Yang,
  Applications of fungal mycelium-based functional biomaterials, in: Fungal
  Biopolymers and Biocomposites: Prospects and Avenues, Springer, 2022, pp.
  147--168.

\bibitem{gandia2021flexible}
A.~Gandia, J.~G. van~den Brandhof, F.~V. Appels, M.~P. Jones, Flexible fungal
  materials: shaping the future, Trends in Biotechnology 39~(12) (2021)
  1321--1331.

\bibitem{adamatzky2021towards}
A.~Adamatzky, A.~Gandia, A.~Chiolerio, Towards fungal sensing skin, Fungal
  biology and biotechnology 8~(1) (2021) 1--7.

\bibitem{adamatzky2021reactive}
A.~Adamatzky, A.~Nikolaidou, A.~Gandia, A.~Chiolerio, M.~M. Dehshibi, Reactive
  fungal wearable, Biosystems 199 (2021) 104304.

\bibitem{chiolerio2022living}
A.~Chiolerio, M.~M. Dehshibi, D.~Manfredi, A.~Adamatzky, Living wearables:
  Bacterial reactive glove, Biosystems (2022) 104691.

\bibitem{nikolaidou2022reactive}
A.~Nikolaidou, N.~Phllips, M.-A. Tsompanas, A.~Adamatzky, Reactive fungal
  insoles, bioRxiv (2022).

\bibitem{whitaker2018electronics}
J.~C. Whitaker, The electronics handbook, Crc Press, 2018.

\bibitem{wilamowski2018fundamentals}
B.~M. Wilamowski, J.~D. Irwin, Fundamentals of industrial electronics, CRC
  Press, 2018.

\bibitem{maini2018handbook}
A.~K. Maini, Handbook of defence electronics and optronics: fundamentals,
  technologies and systems, John Wiley \& Sons, 2018.

\bibitem{jillek2005embedded}
W.~Jillek, W.~Yung, Embedded components in printed circuit boards: a processing
  technology review, The International Journal of Advanced Manufacturing
  Technology 25 (2005) 350--360.

\bibitem{zheng2011review}
L.~Zheng, C.~Y. Wang, Y.~X. Song, L.~Yang, Y.~Qu, P.~Ma, L.~Fu, A review on
  drilling printed circuit boards, Advanced Materials Research 188 (2011)
  441--449.

\bibitem{mumby1989overview}
S.~J. Mumby, An overview of laminate materials with enhanced dielectric
  properties, Journal of Electronic Materials 18~(2) (1989) 241--250.

\bibitem{ehrler2002properties}
S.~Ehrler, Properties of new printed circuit board base materials, Circuit
  World (2002).

\bibitem{mumby1989dielectric}
S.~J. Mumby, J.~Yuan, Dielectric properties of fr-4 laminates as a function of
  thickness and the electrical frequency of the measurement, Journal of
  Electronic Materials 18 (1989) 287--292.

\bibitem{djordjevic2001wideband}
A.~R. Djordjevic, R.~M. Bilji{\'e}, V.~D. Likar-Smiljanic, T.~K. Sarkar,
  Wideband frequency-domain characterization of fr-4 and time-domain causality,
  IEEE Transactions on electromagnetic compatibility 43~(4) (2001) 662--667.

\bibitem{liu2022printed}
H.~Liu, Z.~Gu, Q.~Zhao, S.~Li, X.~Ding, X.~Xiao, G.~Xiu, Printed circuit board
  integrated wearable ion-selective electrode with potential treatment for
  highly repeatable sweat monitoring, Sensors and Actuators B: Chemical 355
  (2022) 131102.

\bibitem{kao2018skinwire}
H.-L.~C. Kao, A.~Bedri, K.~Lyons, Skinwire: Fabricating a self-contained
  on-skin pcb for the hand, Proceedings of the ACM on Interactive, Mobile,
  Wearable and Ubiquitous Technologies 2~(3) (2018) 1--23.

\bibitem{tao2017make}
X.~Tao, V.~Koncar, T.-H. Huang, C.-L. Shen, Y.-C. Ko, G.-T. Jou, How to make
  reliable, washable, and wearable textronic devices, Sensors 17~(4) (2017)
  673.

\bibitem{vieroth2009stretchable}
R.~Vieroth, T.~Loher, M.~Seckel, C.~Dils, C.~Kallmayer, A.~Ostmann, H.~Reichl,
  Stretchable circuit board technology and application, in: 2009 International
  Symposium on Wearable Computers, IEEE, 2009, pp. 33--36.

\bibitem{buechley2009fabric}
L.~Buechley, M.~Eisenberg, Fabric pcbs, electronic sequins, and socket buttons:
  techniques for e-textile craft, Personal and Ubiquitous Computing 13 (2009)
  133--150.

\bibitem{stoppa2014}
M.~Stoppa, A.~Chiolerio, Wearable electronics and smart textiles: a critical
  review, Sensors 14 (2014) 11957--11992.

\bibitem{aerosol}
Optomec, Aerosol jet 200,
  \url{chrome-extension://efaidnbmnnnibpcajpcglclefindmkaj/https://www.optomec.com/wp-content/uploads/2014/08/AJ_200_WEB_0216.pdf},
  [Online; accessed 10-Jan-2023].

\bibitem{heraeus}
H.~Clevios, P jet n v2,
  \url{https://www.heraeus.com/en/hep/products_hep/clevios/clevios_prod/clevios_1.html},
  [Online; accessed 10-Jan-2023].

\bibitem{palmsens}
P.~BV, Palmsens4, \url{https://www.palmsens.com/product/palmsens4/}, [Online;
  accessed 10-Jan-2023].

\bibitem{S5}
Ultimaker, Ultimaker s5,
  \url{https://ultimaker.com/3d-printers/ultimaker-s5-pro-bundle}, [Online;
  accessed 10-Jan-2023].

\bibitem{filaflex}
3DJAKE, Conductive filaflex black,
  \url{https://www.3djake.uk/recreus/conductive-filaflex-black}, [Online;
  accessed 10-Jan-2023].

\bibitem{electrifi}
MULTI3D, Electrifi conductive filament,
  \url{https://www.multi3dllc.com/product/electrifi/}, [Online; accessed
  10-Jan-2023].

\bibitem{bareconductive}
B.~C. Ltd, Bare conductive,
  \url{https://www.bareconductive.com/collections/electric-paint}, [Online;
  accessed 10-Jan-2023].

\bibitem{silver}
XeredEx, Xd-120,
  \url{https://shopee.co.id/XD120-Conductive-Silver-Glue-Wire-Electrically-Paste-Adhesive-Paint-PCB-Repair-i.28090589.2478654570},
  [Online; accessed 10-Jan-2023].

\bibitem{Legend}
EpilogLaser, Legend 36ext tech specs,
  \url{https://www.epiloglaser.com/laser-machines/l36ext-techspecs.htm},
  [Online; accessed 10-Jan-2023].

\bibitem{Tarabella_2020}
G.~Tarabella, D.~Vurro, S.~Lai, P.~D’Angelo, L.~Ascari, S.~Iannotta, Aerosol
  jet printing of pedot:pss for large area flexible electronics, Flexible and
  Printed Electronics 5~(1) (2020) 014005.

\bibitem{secor2018principles}
E.~B. Secor, Principles of aerosol jet printing, Flexible and Printed
  Electronics 3~(3) (2018) 035002.

\bibitem{wilkinson2019review}
N.~Wilkinson, M.~Smith, R.~Kay, R.~Harris, A review of aerosol jet printing—a
  non-traditional hybrid process for micro-manufacturing, The International
  Journal of Advanced Manufacturing Technology 105~(11) (2019) 4599--4619.

\bibitem{mette2007metal}
A.~Mette, P.~Richter, M.~H{\"o}rteis, S.~Glunz, Metal aerosol jet printing for
  solar cell metallization, Progress in Photovoltaics: Research and
  Applications 15~(7) (2007) 621--627.

\bibitem{scalisi2015}
R.~Scalisi, M.~Paleari, A.~Favetto, M.~Stoppa, P.~Ariano, P.~Pandolfi,
  A.~Chiolerio, Inkjet printed flexible electrodes for surface
  electromyography, Organic Electronics 18 (2015) 89--94.

\bibitem{paintdata}
B.~C. Ltd,
  \href{https://cdn.shopify.com/s/files/1/0520/3669/8292/files/EP_tech_data_sheet_a13f3d46-56ce-4689-97cb-da3b3b3d52d2.pdf?v=1655713221}{Electric
  paint. technical data sheet}, last accessed 13 December 2022 (2017).
\newline\urlprefix\url{https://cdn.shopify.com/s/files/1/0520/3669/8292/files/EP_tech_data_sheet_a13f3d46-56ce-4689-97cb-da3b3b3d52d2.pdf?v=1655713221}

\bibitem{CW2400}
Chemtronics, Chemtronics cw2400 liquid adhesive,
  \url{https://uk.rs-online.com/web/p/adhesives/0496265}, [Online; accessed
  10-Jan-2023].

\end{thebibliography}

\end{document}